\documentclass[iop]{emulateapj}
\usepackage{epstopdf}
\usepackage{graphicx} 
\usepackage{amsmath} 
\usepackage{rotating}
\usepackage{amssymb}

\usepackage{color}
\usepackage{ulem}
{

\newcommand{\changes}[1]{}

\def\wig#1{\mathrel{\hbox{\hbox to 0pt{%
          \lower.5ex\hbox{$\sim$}\hss}\raise.4ex\hbox{$#1$}}}}

\shorttitle{Circumplanetary Disks around Wide-Orbit Planets}
\shortauthors{Shabram, et al.}

\newcommand{\mj}{$M_{\mathrm{J}}$}

\newcommand{\msun}{$M_{\odot}$}

\newcommand{\cp}{\citep}
\newcommand{\ct}{\citet}

\slugcomment{}

\begin{document}

\title{The Evolution of Circumplanetary Disks around Planets in Wide Orbits: Implications for Formation Theory, Observations, and Moon Systems}

\author{Megan Shabram\altaffilmark{1}, Aaron C. Boley\altaffilmark{1}$^{,}$\altaffilmark{2}} 

\altaffiltext{1}{Department of Astronomy, University of Florida, 211 Bryant Space Center, Gainesville, FL 32611}
\altaffiltext{2}{Sagan Research Fellow}

\begin{abstract} 
Using radiation hydrodynamics simulations, we explore the evolution of circumplanetary disks around wide-orbit proto-gas giants.  
At large distances from the star ($\sim100$ AU),  gravitational instability followed by disk fragmentation can form low-mass substellar companions (massive gas giants and/or brown dwarfs) that are likely to host large disks.  
We examine the initial evolution of these subdisks and their role in regulating the growth of their substellar companions, as well as explore consequences of their interactions with circumstellar material.
We find that subdisks that form in the context of GIs evolve quickly from a very massive state.  Long-term accretion rates from the subdisk onto the proto-gas giant reach $\sim0.3$ Jupiter masses per kyr.
We also find consistency with previous simulations, demonstrating that subdisks are truncated at $\sim1/3$ of the companion's Hill radius and  are thick, with $(h/r)$ of $\gtrsim0.2$.  The thickness of subdisks draws to question the use of thin-disk approximations for understanding the behavior of subdisks, and the morphology of subdisks has implications for the formation and extent of satellite systems. 
These subdisks create heating events in otherwise cold regions of the circumstellar disk, and serve as planet formation beacons that can be detected by instruments such as ALMA.
\end{abstract}


\section{Introduction}
At orbital radii $\gtrsim100$ AU, gravitational instability (GI) may be the principal formation mechanism for massive planets and brown dwarfs \cp{Stamatellos07, Boley09}.  
GIs can occur in self-gravitating disks whenever the destabilizing effects of gravity become comparable to the stabilizing effects of disk pressure and shear \cp{Toomre64}.  
These instabilities typically manifest themselves as spiral structure \cp{Durisen07}, which result in disk heating and mass redistribution through shocks and gravitational torques.
In cases of very strong instability, spiral arms are unable to self-regulate and these arms can fragment into bound clumps. 
Because fragmentation is most likely to occur at distances $\sim 100 $ AU \cp{Rafikov09, Boley09, Clarke09}, nascent fragments are a few AU in radius and contain mostly molecular gas.  
Their large size makes these clumps susceptible to strong tidal effects, including complete destruction due to inward migration or due to pericenter passages if on eccentric orbits \cp{Boley10,Nayakshin10}.
As fragments radiate away energy, they contract, leading to a rise in internal temperature.
Eventually, temperatures become high enough to cause $\mathrm{H}_2$ dissociation, which leads to a rapid collapse of the clump as energy goes into dissociation instead of increasing the gas pressure.
At this stage, the low-angular momentum material will form a protoplanet or proto-brown dwarf that is no longer susceptible to tidal stripping by the host star.  
The high-angular momentum material will be unable to fall directly on the core's surface, and an accretion disk will form.
In this study, we focus on the initial evolution of a circumplanetary/brown dwarf disk (forthwith subdisk) and its evolution within a circumstellar disk in the context of an object formed by GIs.   

Characterizing the effects of a high-mass ratio binary (star/planet) on subdisk evolution is a necessary step toward understanding the growth of massive planets.
Circumplanetary disk evolution is also of interest for satellite formation, as subdisk structure, lifetimes, and size will impact the formation of large moons.
A combination of analytics \cp{Quillen98, Canup02, Ward10, Martin11} and simulations \cp{D'Angelo02, Machida08, Ayliffe09, Martin11} have been used to explore the morphology and evolution of circumplanetary disks.  General features that have been found include subdisk truncation at $\sim$ 1/3 of the host planet's Hill radius  and  complex, three-dimensional gas flows.    

Most studies to date have focused on low-mass planets, and few of these studies use three-dimensional hydrodynamics \cp[see][for examples of simulations in 3D]{Machida08, Ayliffe09}.  Moreover, the inclusion of radiative transfer in these simulations is rare, which can play a significant role in regulating the evolution of the system \cp{Ayliffe09, Ayliffe11}.  In this study, we focus on  the early evolution of newly-formed massive subdisks that are expected to form during massive planet and brown dwarf formation through disk fragmentation. 

This work uses three-dimensional radiation hydrodynamics simulations  to explore the evolution of subdisks embedded in circumpstellar disks.  The simulations are designed to address the following questions: (1) Is self-gravity likely to play a role in the evolution of subdisks born by disk fragmentation?  (2) What is the role of the subdisks in regulating the mass growth of the proto-gas giant or proto-brown dwarf? (3) What are the observable signatures of nascent, wide-orbit proto-gas giants? 

This manuscript is organized as follows.
In \textsection 2, we describe the computational method behind our radiation hydrodynamics simulations.  The results are presented in \textsection 3, and their implications are discussed in \textsection 4 along with our conclusions.
 
\section{Computational Method}
We aim to simulate and characterize the initial evolution of subdisks around newly-formed GI protoplanets.  
Simulations are run using CHYMERA, an Eulerian code that solves the equations of hydrodynamics on a fixed, cylindrical grid \cp{Boley07}. 
The subdisks are evolved using $r, \phi, z = 256,\;512,\;64$ cells centered on the host planet.  
Each subdisk has a resolution of $\Delta r, \Delta \phi, \Delta z = 0.1\,\mathrm{AU},\, 2\pi\, r/512,\, 0.1\,\mathrm{AU}$.  
First, we investigate an isolated subdisk (no circumstellar disk), referred to as simulation (a), which is used for a base for comparisons. 
We then explore effects of a circumstellar disk on the evolution of the subdisk by allowing mass to flow onto the computational grid.  This mass flux is varied  with a gas volume density of $1 \times 10^{-14}$, $3 \times 10^{-14}$ and $1 \times 10^{-13}$ $g/cm^3$ for simulations (b), (c) and (d), respectively.   \changes{We choose these values for the volume densities based on global simulations of fragmented disks from \ct{Boley09} and \ct{Boley10} and to explore a range of possibilities, including low- and high-density interarm regions. } Further details are given in section \ref{sec:massflow}. 

Our simulations are initialized with a subdisk-to-planet mass ratio $\sim 1$ (3 \mj{}  each). 
The proto-gas giant and circumplanetary disk orbit the 1 \msun \;primary at 100 AU. 
The initial conditions (ICs) for the simulations are created using analytic approximations for the disk (axisymmetric, Keplerian rotation about the proto-gas giant, and non-self-gravitating vertical structure with a constant adiabatic index).  Using these approximations the surface density is tuned to give an initially flat Toomre Q radial profile \cp{Toomre64} with $Q\sim 1.1$. 
When $Q\lesssim 1.7$, disks are susceptible to the formation of spiral arms (e.g., \ct{Durisen07}) and $Q\sim 1$ is a very unstable disk.  
When the disk is run in complete isolation (no external potential from the host star), the disk is overall stable, although the approximations lead to vertical and radial oscillations.
These out-of-equilibrium features are minor compared with the impact of including the host star's potential.  
For this reason, quieter ICs are not developed, as the tidal potential (section \ref{sec:tides}) quickly alters the initial setup.

Self-gravity, radiation transport, and the tidal potential from the primary are included in our simulations.  
We use a modified version of the BDNL scheme from \ct{Boley07} to  handle the cooling of the gas.  
The algorithm invokes flux-limited diffusion (FLD) in the radial and azimuthal directions, while  
ray tracing is used for the vertical direction where the majority of disk cooling occurs due to the large surface area of the photosphere.   
At very low optical depths, radiative transfer is in the free-streaming limit and the material cools slowly, as gas becomes inefficient at
emitting and absorbing radiation.  At very large optical depths, the high opacity also makes cooling inefficient.  
Ray tracing ensures that the transition region between these two limits, which is where most of the disk cooling occurs, is followed accurately.
The equation of state from \ct{Boley07b} is used, with the ortho-para hydrogen ratio fixed to 3:1.  For the temperatures explored here, the gas is well approximated by an adiabatic index of 5/3.

\subsection{Modifications} \label{2.1}
\subsubsection{Radiative Transfer \label{eqn:radiative_transfer}}
We use a subcycling method to control heating and cooling. 
When the radiative transport routine is called, we first calculate the divergence of the flux for each cell. 
The ratio of the local internal energy density to the local divergence of the flux gives a cell-wise radiative time step. 
We then compare the shortest radiative time step over all cells with the Courant time step. 
Ideally, the shortest radiative time step must be resolved in order to keep the algorithm stable.  
Time steps that are too large can produce excessive heating or cooling of the gas, leading to unphysical results and numerical instability. 
We address this issue by allowing the radiative transfer to subcycle in cases where the radiative timescale is shorter than the hydrodynamic timescale, i.e., allow multiple calls to the radiative transfer routine for any one hydro step.
We apply the divergence of the flux to the radiative time step in order to update the internal energy density for each subcycle. 
If the cooling/heating time has not reached the hydrodynamic timescale after 8 subcycles, we allow the gas to cool to no lower than the background temperature (see section 2.3).
We also limit the heating for the last subcycle to a 10\% increase in internal energy density. 
We calculate the optical depth once per cycle holding the value constant during the iterative cooling routine.
This method provides numerical stability and reasonable speed while still attempting to resolve the actual radiative time scale.

As subdisks are not isolated, they will be affected by incident radiation from their environment.  We include this effect by setting a minimum irradiation background temperature, as a function of distance from the protoplanet.  This reflects the temperature required to balance the radiation received by an otherwise cold subdisk:
\begin{equation}
T_{irr}^4 = \left( \frac{1}{2} \right) \left( T_e^4 + T_{acc}^4 \right) \left( \frac{R_p}{D} \right)^{3} + T_B^4.
\end{equation}
Here, $R_p$ is the host planet radius, which we set to 2 $R_J$, $D$ is the radial distance from the subdisk center,  $T_e$ is the effective temperature of the host planet, which we take to be $1000K$, and $T_B$ is the ambient background temperature due to the star and its environment, which we set to $10K$.  \changes{These values represent a working guess for conditions of a newly-formed protoplanet \citep[e.g.,][]{spiegel_burrows_2012} and background temperature at large disk radii. } The distance dependence is based on a thin, irradiated disk.  
$T_{acc}$ is the temperature corresponding to the luminosity that results from mass accretion from the subdisk onto the protoplanet, and is described in more detail in section \ref{sec:massflow}.

\subsubsection{Tidal and Indirect Potentials\label{sec:tides}} 
Simulations are performed in the frame of the planet, with the planet fixed to the center of the computational grid. The star's position is fixed in this reference frame at $\vec{x}_\star=x_\star \hat{e}_x=-100$ AU.
To accommodate these choices,  modifications are made to the potential field and the equations of motion.  
Let the total potential be
\begin{equation}
\Phi_{tot} = \Phi_{\star} + \Phi_{p} + \Phi_{ind} +  \Phi_{sg} +\Phi_{R}.
\end{equation}
Here, the potential components from the star and planet are written as $\Phi_\star$ and $\Phi_p$, respectively.  The indirect potential, $\Phi_{ind}$, captures the subdisk's influence on the planet, even though the planet is held fixed to the grid center \citep[see][]{Michael10}.  The self-gravity of the gas is given by $\Phi_{sg}$, and the rotational potential,
\begin{equation}
\Phi_{R} =-\frac{1}{2}\Omega^{2}|\hat{e}_z\times(\vec{x} - \vec{x_{\star}})|^{2},
\end{equation}
takes into account the star-planet orbit.  The vector $\vec{x}$ is the position of a grid cell relative to the planet.  Because the mass on the grid can be variable, we hold $\Omega^2 = G(M_{\star}+M_p)/x_\star^3$ fixed throughout the simulations.  By including $\Phi_R$ in the total potential, the equations of motion do not need to be changed to include the  centrifugal terms.  

The Coriolis effect is not implicitly included in the potential terms, and must be added to the force calculation.  Let the radial momentum and angular momentum densities be represented by $S$ and $A$, respectively.  The Coriolis terms on our cylindrical grid are then
\begin{equation}
S = S + 2 \Omega v_{\phi} \rho \Delta t ,
\end{equation}
\begin{equation}
A = A - 2 \Omega v_{r} r \rho \Delta t  
\end{equation}
for azimuthal and radial velocities $v_\phi$ and $v_r$, respectively, gas mass volume density $\rho$, and cylindrical subdisk radius $r$.
With these updates applied during sourcing, the calculation can now be performed in the frame of the planet. 

\subsubsection{Mass Flow Boundaries \label{sec:massflow}}
The disk boundary conditions are handled using both inflow and outflow boundary cells.   
When material flows through the inner grid boundary (0.8 AU), mass is accreted by the protoplanet and included in the subsequent potential calculations. 
The process of accretion should generate an additional luminosity source that will irradiate the subdisk.  
We account for this effect by modifying our incident temperature profile, which sets the background radiation field at a given distance from the planet.  
This luminosity can be described by its temperature as
\begin{equation}
T^4_{acc} = \frac{GM_{p}\dot{M}}{8\pi \sigma R_p^3} ,\label{eqn:t_acc}
\end{equation}
where $\sigma$ is the Stefan-Boltzmann constant.  For each call to the radiative transfer routine, $T_{acc}$ is calculated based on the current mass accretion from the subdisk $\dot{M}$.  This value is then used to set the background temperature as described in section \ref{eqn:radiative_transfer}.

The outer radial boundary uses outflow conditions unless the given boundary cell is set to flux mass onto the computational grid.  In the latter case, mass flux is envisaged to be part of the primary disk.  One simulation (sim a) does not include any mass fluxing, so all radial boundary cells use outflow conditions.  Simulations (b), (c), and (d) are run with incoming gas volume densities set to $1\times 10^{-14}$, $3 \times 10^{-14},$ and $1 \times 10^{-13}$ $g/cm^3$, respectively.  \changes{We set up a Keplerian shear over the entire grid, where gas is fluxed onto the grid between azimuths of (0,$\pi/2$) and ($\pi$,$3\pi/2$) with a velocity determined by the difference between the Keplerian azimuthal speeds of the protoplanet and gas at a given location.  The resulting velocity vectors that trace the flow are similar to that seen in  global fragmentation models.}  For simplicity and overall stability, gas is fluxed only up to half the vertical component of the domain. 
The disk's upper boundary is only an outflow boundary.
As will be shown below, the chosen volume densities lead to plausible surface densities for a fragmented protoplanetary disk at 100 AU.

\section{Results}
\subsection{Mass Evolution}
\begin{figure}
\centering
\includegraphics[width=0.8\textwidth,natwidth=610,natheight=642]{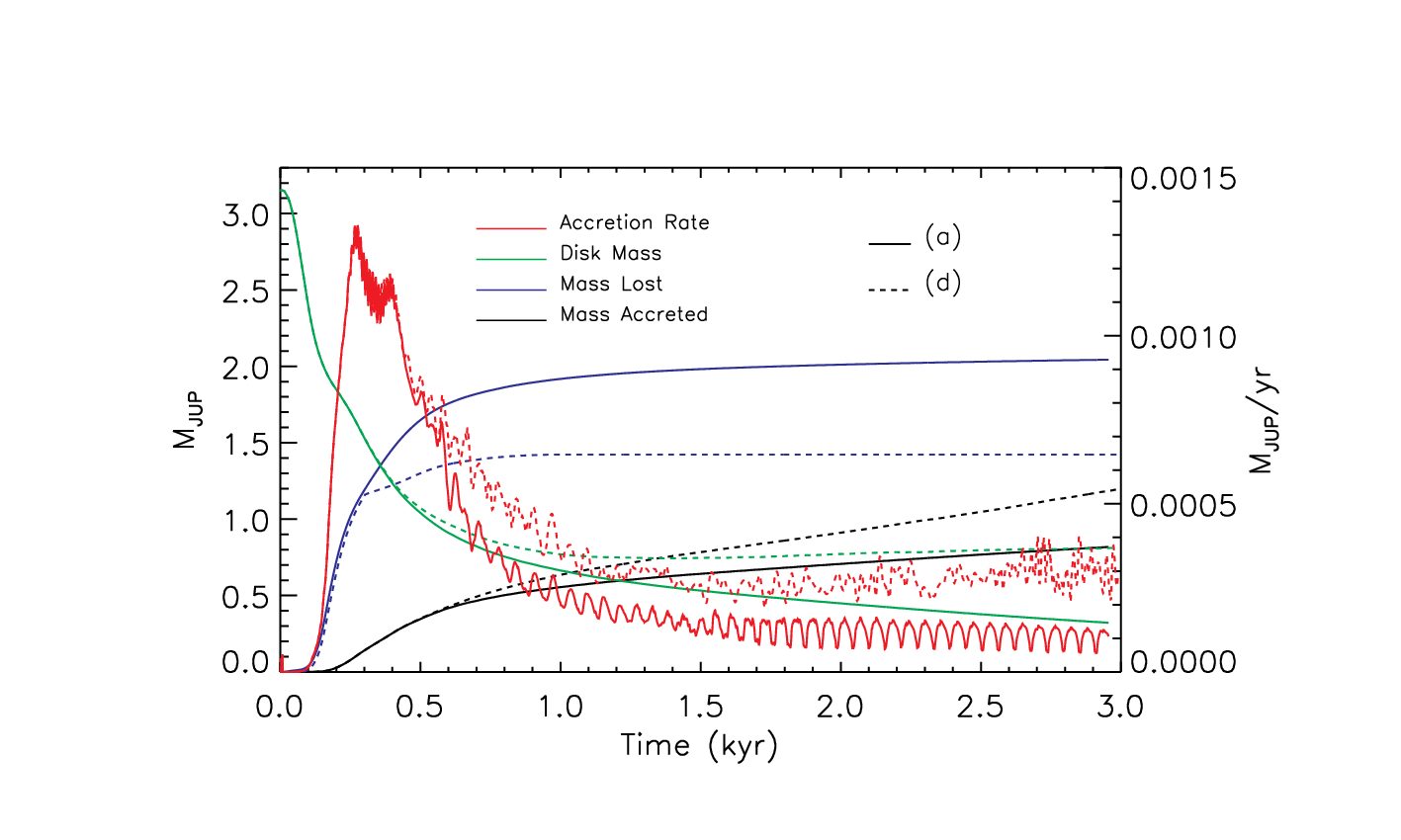}
\caption{Mass accretion history for the planet and subdisk for simulations (a) (no mass flux) shown as solid lines and simulation (d) (high mass flux) shown as dotted lines (see Table 1).  
The curves show the evolution of the subddisk mass (red), the mass lost from the computational grid (blue, isolated case only), the mass accreted by the planet (black), and the accretion rate onto the planet in $M_J\,\rm yr^{-1}$. 
We find that all subdisks evolve away from their initial conditions within one orbit of the protoplanet about the star (1 kyr), which corresponds to about 7 orbits within the subdisk at 5 AU from the protoplanet (about 1/2 the Hill radius). 
Most of the mass is accreted during this time period, reaching about 0.5 \mj\, for all simulations.  
After about 1.5 kyr, the mass accretion rate settles down to a roughly steady state, with $\sim 0.29$ \mj~per kyr in simulation (d), 0.14 in simulation (c), 0.11 in simulation (b), and 0.10 in simulation (a).  
The subdisk in simulation (a) is not being continuously fed, so this rate can only persist for another $\sim 6$ kyrs.\label{fig:mass_evol}}
\end{figure}

We explore subdisk evolution by varying the amount of material that is allowed to flow through the Hill region in 4 different simulations (see Table \ref{table1}). 
The results show that the interactions between the subdisks and the circumstellar disks will have a large influence on the host planet.  
Figure 1 highlights this evolution for simulations (a) and (d), which shows the rate of mass flowing from the subgrid onto the protoplanet (specifically the mass flowing through 0.8 AU), the total mass on the computational grid, and the cumulative mass that leaves the computational domain other than through accretion by the protoplanet. 
This mass lost profile shows an abrupt kink in simulation (d) and levels off soon after.  
This is a result of mass being fluxed onto the grid, which quickly balances the mass that is lost.  
We find that all subdisks evolve away from their initial conditions within one orbit of the protoplanet about the star (1 kyr), which corresponds to about 7 orbits within the subdisk at 5 AU from the protoplanet (a little less than 1/2 the Hill radius). 
Most of the mass is accreted during this time period, reaching about 0.5 \mj\, for all simulations.  
After about 1.5 kyr, the mass accretion rate settles down to a roughly steady state (Fig.~\ref{fig:mass_evol}), with $\sim0.29$ \mj/kyr in simulation (d), 0.14 in simulation (c), 0.11 in simulation (b), and 0.10 in simulation (a).  
The subdisk in simulation (a) is not being continuously fed, so this rate can only persist for another $\sim 6$ kyrs.  

Simulation (d) retains the most massive subdisk, but even this disk quickly evolves away from its highly unstable state.  The minimum $Q = c_s \kappa/(\pi G \Sigma)$, for epicyclic frequency $\kappa$, at the end of the simulation is $\sim 1.7$ for a narrow region at a subdisk radius $\sim 2.5$ AU.  While this region is at the cusp of forming non-axisymmetric structure due to GIs, structure appears to be dominated by the tidal potential, i.e., two distinct spiral arms aligned with the tidal distortion.

\subsection{Morphology}
Figures 3 and 4 show the end-state midplane temperature and surface density images for each simulation.  To define the limits of the disk, we also show in Figure 2 the cumulative mass profile as a function of distance from the protoplanet.  In all simulations, the subdisk is truncated abruptly at $\sim 4$ AU, which is approximate 1/3 of the system's Hill radius, consistent with previous work.  This truncation can be seen in the surface density images as well, and is highlighted by the circle in the midplane temperature images.  The surface density images also show spiral arms in the subdisk resulting from the tidal potential, as well as the spiral wakes, which penetrate from the circumstellar region all the way to the subdisk.  Material enters the Hill sphere over a wide range of azimuths and creates shocks as circumstellar material meets the subdisk.    

The interactions between the circumstellar disk and gas within the Hill region lead to shock structures that outline the wakes and circumstellar-subdisk boundary.  These features are most pronounced in simulations (c) and (d). Figure 5 further highlights this, showing the velocity flow of the gas around the Hill region in simulation (d). The heating of this gas could produce molecular walls as icy dust passes through the Hill sphere.  While the subdisk itself also provides a region for heating ices, the molecular walls provide an additional signature of an embedded clump.   We highlight this in two different ways. First, we show in Figure 6 a radial cut at about 6 o'clock for each panel in Figure 3.  There is a rapid dropoff to the background temperature, followed by an isolated temperature spike in the Hill sphere shock region, i.e., the potential molecular wall.  While this region is still fairly cold, it will affect some ices such as CO.  Moreover, we have only explored one location in the circumstellar disk in these simulations, and similar features may be present at higher temperatures when the protoplanet is placed at smaller orbital radii for analogous simulatiuons.  

\changes{Next, we show in Figure 7 the midplane abundance of CO in the gas phase relative to a total CO abundance of $n_{\rm CO}=10^{-4} n_H$, where $n_H$ is the total number of hydrogen nuclei and $10^{-4}$ represents the typical ISM CO abundance relative to hydrogen.    To calculate the abundance of CO in the gas phase, we use a simple chemical equilibrium model. 
Let $R_a = \sigma_d v n_{\rm CO} n_d$ be the absorption rate of CO onto dust grains, where $\sigma_d$ is the cross-sectional area of the dust, $v= (8 k_B T /[\pi m_{\rm CO}])^{1/2}$ is the speed of a CO molecule based on gas temperature $T$ and CO mass $m_{\rm CO}$, and $n_d$ is the number density of dust.  The desorption rate is given by $R_e=\nu_0 \exp(-B/T)$, where $\nu_0 \approx 1000$ GHz and $B=1000$ K \cp{watson_salpeter_1972,Hasegawa92,brinch-etal-2008}.  The gas-phase CO number density is then given by $n_{\rm CO} -R_a/R_e$. 
We use typical ISM dust properties for $n_d\sim 2.6\times10^{-12} n_H$ and $\sigma_d\sim3\times10^{-10} {\rm cm}^{-2}$ \cp{Hasegawa92}. 
Gas temperature and densities are derived directly from the simulations, and the above analysis is used to determine CO column density and fractional gas-phase abundances for the end snapshot of simulation (d) (Fig.~7). 
We have compared the derived rates with cell-crossing times of the gas, and gas-grain exchanges are typically more rapid than a cell crossing, indicating that the equilibrium condition used to create Figure 7 is reasonable for this case.  Material that is shocked as it falls into the Hill region offers a location for gas-phase chemistry that would otherwise not exist, and has the potential to create observational signatures, as described in the next section.} 

\subsection{Prospects for ALMA} 
In the previous section, we discussed how protoplanets/brown dwarfs in the outer regions of young, circumplanetary disks will alter the thermodynamics and the chemistry of the surrounding gas.  
Even if the protoplanet alone is difficult to detect, a relatively hot subdisk could significantly increase the continuum and line emission in localized regions of circumstellar disks.  Morphological structures such as molecular walls, as defined above, can further reveal the presence of an embedded object.    As shown in Figure 7, gas-phase CO column densities are $\sim10^{18}$ cm$^{-2}$ in the molecular walls  and $\sim10^{23}$ cm$^{-2}$ in the subdisk.  Not only could the resulting emission be detected, but it could also be resolved. Molecular wall separations from the molecular subdisk are $\sim$ few to 10 AU wide, requiring at least 0.\arcsec1 resolution.  


\changes{To evaluate the typical flux densities that we could expect in simulation (d), consider CO line emission. Due to the gas densities, we assume that the excitation temperature for CO and the gas temperature are the same. 
 With this approximation, the populations of rotational states among the CO molecules is given by 
\begin{equation}
\frac{n_{J+1}}{n_J}=\frac{2J+3}{2J+1}\exp[-2(J+1)hB/(kT)]
\end{equation}
for lower energy state $J$, Planck and Boltzmann constants $h$ and $k$, and rotational constant $B=57.64$ GHz\citep{splatalogue}. At 25 K, we find that $\sim 30$\% of the CO gas is in the $J+1=2$ state. 
We can expect considerable emission from the  CO at the CO($2\rightarrow1$) transition.  
However, before the expected flux density can be determined, we need to estimate whether the line will be optically thin or thick.  
Following \cite{caselli-etal-2002}, we write the total CO column density, as a function of line optical depth, as
\begin{eqnarray}
N_{{\rm CO},\tau} & = & \frac{8 \pi (2(J+1)B)^2}{c^2 A(CO)_{J+1,J}\phi(\nu)}\\\nonumber
              &   \times &\frac{2J+1}{2J+3}\frac{\tau }{(1-\exp[-2(J+1)hB/(kT)])}\\
              & \times & \frac{Q_{\rm rot}}{(2J+1)\exp[-J(J+1)hB/(kT)]}\nonumber
\end{eqnarray}
where $A(CO)_{J+1,J}$ is the Einstein emission coefficient, 
\begin{equation}
Q_{\rm rot} = \Sigma_{J=0}^{\infty} (2J+1)\exp{[-J(J+1)hB/(kT)]}
\end{equation}
 is the rotational partition function for a molecule that has a simple rotational ladder, such as CO, and $\phi(\nu)$ is the line profile in $\rm Hz^{-1}$, which we take to be thermal.  }

\changes{For the CO(2$\rightarrow 1$) transition, i.e., $J=1$ and $A(CO)_{2,1}\sim7\times10^{-7}$  Hz \citep{splatalogue}, the $\tau=1$ surface will be reached at a total CO column density of $\sim2\times10^{15}$ cm$^{-2}$.  The average column density through the molecular walls is $\sim 4\times10^{18}$ cm$^{-2}$, making the line emission optically thick.  In this case, the flux density is simply
\begin{equation}
S_\nu = B_\nu(T) ({\rm wall~area})/{D^2},
\end{equation}
where the {\it wall area} is the total area of the molecular walls along the line of sight, $D$ is the distance to the source, and $B_\nu$ is the Planck function.  At $T=25$ K and $D=100$ pc, $S_\nu \sim 8$ mJy at 230 GHz.  The emission is easily detectable by ALMA, making the instrument a fundamental tool for investigating planet/brown dwarf formation at large distances.  A combination of compact and extended configurations can be used to first detect the emission and then to look for structure.  }

\subsection{Vertical Structure: Thin vs Thick Circumplantery disks}
Figure 8 shows the end-state $h/r$ for simulation (d), where the disk scale height $h(r)=c_s(r)/\Omega(r)$. 
The sound speed $c_s$ and orbital frequency $\Omega$ at a given $r$ are determined using mass-weighted averages in the azimuthal and vertical directions. 
For the simulated disk, the $h/r\sim0.25$ at the $r\sim 1$ AU and $0.5$ at $r\sim 5$ AU.
At even larger radii, $h/r\sim1$, demonstrating that material flowing into the Hill sphere is not initially part of a disk, giving rise to complex three-dimensional structure.  
This is highlighted by a meridional slice through the substellar and antistellar locations (Fig.~9). 

These subdisks are thick ($h/r\gtrsim0.2$), and may not be well-approximated by thin-disk theory ($h/r\lesssim 0.1$).  
The disparate concentrations of gas mass between subdisks and circumstellar disks have implications for understanding moon formation, as dust and ice coagulation, settling, and moon migration  could be very different in the thick disk regime.  These processes have largely been explored in the context of thin-disk theory (e.g., Canup \& Ward 2002), so simulations of gas-solid coupling and satellite migration in the context of disks with large scaleheights should be conducted. 

Why are these subdisks thick?  Even though subdisk temperatures are low, subdisks are very hot compared with their orbital speed ($h/r=c_s/[r\Omega]$).  At the outer boundary of the disk,
the subdisk temperature would need to be 0.4 K for $h/r\sim0.1$, well below the CMB. At a subdisk radius of 1 AU, the disk temperature would need to be less than 10 K (below the stellar irradiation temperature) to have an $h/r\sim 0.1$ for these simulations.  At very small subdisk radii, the $h/r$ may settle toward typical values in the thin-disk regime, as suggested by Figure 8.  Once this regime is reached, then standard accretion disk theory may be applicable, but most of the subdisk mass will likely be contained in the thick-disk region.

Finally, for reference, we calculate the effective $\alpha$ for a Shakura \& Sunyaev (1973) $\alpha$-disk that would be commensurate with the accretion rates seen in simulation (d).  We calculate $\alpha$ using a steady-state solution, i.e., $\alpha\approx \dot{M}/(3\pi c_s^2\Omega^{-1}\Sigma)$.  For this estimate, we use the azimuthal average for values at a subdisk radius $\sim 1$ AU, which yields a value of $\alpha = 0.0035$.    

\section{Discussion and Conclusions}
Planet growth and the formation of moon systems is dependent on the time evolution of circumplanetary disks.  
Radiation hydrodynamic simulations of these systems in early stages of formation can give insight into the detailed physics of these processes.
In this work, we have used radiation hydrodynamics simulations to investigate the initial evolution of  circumplanetary disks under four different assumptions for mass inflow from a circumstellar disk.
Despite the subdisks being initialized in an extremely unstable state ($Q\sim 1$), the disks evolve rapidly away from strongly self-gravitating conditions.  
Simulation (d) does show a minimum $Q\sim1.7$ at a subdisk radius 2.5 AU.  
This is at the boundary for the development of spiral distortions, but disk asymmetries appear to be driven by tidal features.
While we cannot make a strong claim about the stability of the subdisk interior to $\sim 1$ AU, the simulations presented here do argue against a sustained strongly gravitationally unstable subdisk, where $Q$ is near fragmentation conditions $Q\lesssim1.4$ (Mayer et al.~2004).  

After the initial phase of rapid evolution, each disk settles toward a rough steady state.  
Tides from the star provide a strong enough perturbation on the subdisk to keep material flowing inward even though the subdisk is no longer significantly self-gravitating.   
The accretion rate of the subdisk onto the protoplanet is dependent on the mass of the circumstellar disk, although this dependency is shallower than linear for the regime explored here. 
In simulation (d), which is the most massive circumstellar disk used in these simulations, the mass accretion from the subdisk onto the planet is $\sim 0.3~M_J$ per kyr.  At this rate, we expect the protoplanet to double its initial mass in about 10 orbits.  It should be noted that GI-infected circumstellar disks can evolve significantly on timescales of $10^4$ yr \cp{Boley06}, so some caution should be used when extrapolating the mass growth beyond this point.  \changes{If, on the other hand, accretion does continue at the level of simulation (d) for tens to a hundred orbits, the protoplanet could grow to become a brown dwarf or even a low mass star, as seen in some global simulations \citep[e.g.,][]{stamatellos_withworth_2009}}.   We also note that an accretion mechanism will need to be available for the interior region of the subdisk ($<0.8$ AU here) for subdisk mass to reach the proto-gas giant/brown dwarf, which is not captured in these simulations. 

Disk truncation is abrupt, and occurs at about 1/3 of the protoplanet's Hill sphere.  The specific angular momentum distribution of material in the disk reaches a peak at about a subdisk radius of 3.5 AU, with a value $J_{\rm max} \sim 6.5\times 10^{17}$ cm$^2$ s$^{-1}$. A common parameterization for this value is often written as $J_{\rm max}=\ell \Omega_\star R_H^2$, where $R_H\sim 12$ AU for these simulations.  The factor $\ell$ is the angular momentum bias, and its value has been of considerable interest for creating circumplanetary disk models (e.g., Ward and Canup 2010).  Based on our simulation-derived values, $\ell\sim 0.6$.  We note that $\Omega$ can deviate considerably from the Keplerian value at the location of $J_{\rm max}$, so $\Omega$ is taken directly from the simulation as done for the disk scale height calculation. 

Tidal effects from the host star on the subdisk, as well as material that falls into the Hill sphere from the circumstellar disk, can produce gas temperature changes that create regions with gas-phase molecules that would otherwise be frozen on dust grains.  These molecular walls, in addition to hotter regions of subdisks, could allow gas-phase processes in regions of circumstellar disks that would otherwise be too cold for certain interactions, as well as provide observational diagnostics. 
For example,  we find CO column densities of $\sim 10^{16}$ to at least $10^{23}$ cm$^{-2}$, which could produce molecular emission observable with ALMA with potentially resolvable morphological structures.  

In addition to asymmetric structures, we find that subdisks have large $h/r$ ratios, consistent with previous work that used very different methods.  The temperature required for these subdisks to be in the thin-disk regime is lower than realistic background temperatures, forcing subdisks to be thick.  Three-dimensional simulations of subdisks that explore dust-solid evolution, including satellite-subdisk interactions, need to be explored in these thick disks. 

The answers to the three principal science questions of this work can be summarized as follows: \changes{(1) These subdisks show a rapid departure from a strongly self-gravitating state.  (2)  The initial evolution of these systems demonstrate that subdisks can deliver a large amount of mass onto the protoplanet if the subdisk becomes heavily mass loaded.  Passage through spiral arms may create a realistic path for driving periods of very rapid evolution and mass accretion through the subdisk. In contrast, the accretion rate of material onto the proto-gas giant at the end of the simulations is dependent on the local mass in the circumstellar disk. There is some indication that the subdisk may play a limited role in regulating the mass flow, as the accretion rate does not scale linearly with mass. In particular, only simulation (d) shows a strong increase in the mass accretion onto the protoplanet at the end of the simulation. Nonetheless, taking into account the rapid evolution of the subdisk during the initial phases of evolution,  any regulation of mass growth by the subdisk may be restricted to low-mass circumestellar disks regions, such as between spiral arms. } (3) Subdisks create regions in outer disks that have higher temperatures than what would normally be present.  These regions can lead large deviations in gas phase molecular abundances, which can be used for observational diagnostics.  Gas-phase chemistry that would normally be frozen out could be enhanced in these regions.

\changes{Finally, exploring the effects of inward migration on subdisk evolution and protoplanet/brown dwarf mass growth is a necessary future step.  During inward migration, the tidal field will gradually increase in strength, causing stronger and stronger perturbations to the existing disk mass.  This alone could increase the accretion rates through the subdisk.  On the other hand, as the protoplanet moves inward, its Hill sphere shrinks, limiting the amount of mass that can move into the Hill sphere.  The relative importance of these effects needs to be addressed.   Furthermore, these simulations assume that protoplanet has already collapsed through H$_2$ dissociation and has become compact. As a result, we do not expect the protoplanet to be fully or partly destroyed through tidal downsizing \citep{Boley10,Nayakshin10}.  However, inward migration has the potential to wreak havoc on any nascent moon system, if moons can indeed form early in such violent environments.}

\acknowledgements
We thank Paola Caselli, Peter Barns, and Eric Ford for discussions that improved this manuscript. 
We thank the referee for very helpful suggestions and comments.
M.S.'s support was provided in part by the National Radio Astronomy Observatory (NRAO) through award SOSPA0-007. 
A.C.B.'s support was provided by a contract with the California
Institute of Technology (Caltech) funded by NASA through
the Sagan Fellowship Program.
\bibliographystyle{apj}


\pagebreak

\begin{table}[ht]
\caption{Initial Conditions}
\centering
\begin{tabular}{ c c c c c}
 All simulations & & &Simulation & Mass flux $(g/cm^3)$\\ \hline
 \hline
 Disk mass & 3 \mj & & (a) & $0$\\ \hline
 Host planet mass & 3 \mj & & (b) & $1 \times 10^{-14}$\\ \hline
 Barycenter of subdisk orbit & $100AU$ & & (c) & $3 \times 10^{-14}$\\ \hline
 Disk radius & $10AU$ & & (d) & $1 \times 10^{-13}$\\ \hline
 Initial Toomre Q & \textasciitilde1.1 & & &\\ \hline
Host star mass & 1\msun & & &\\ \hline
\hline
\end{tabular}
\label{table1}
\end{table}

\begin{figure}
\centering
\includegraphics[width=0.6\textwidth,natwidth=610,natheight=642]{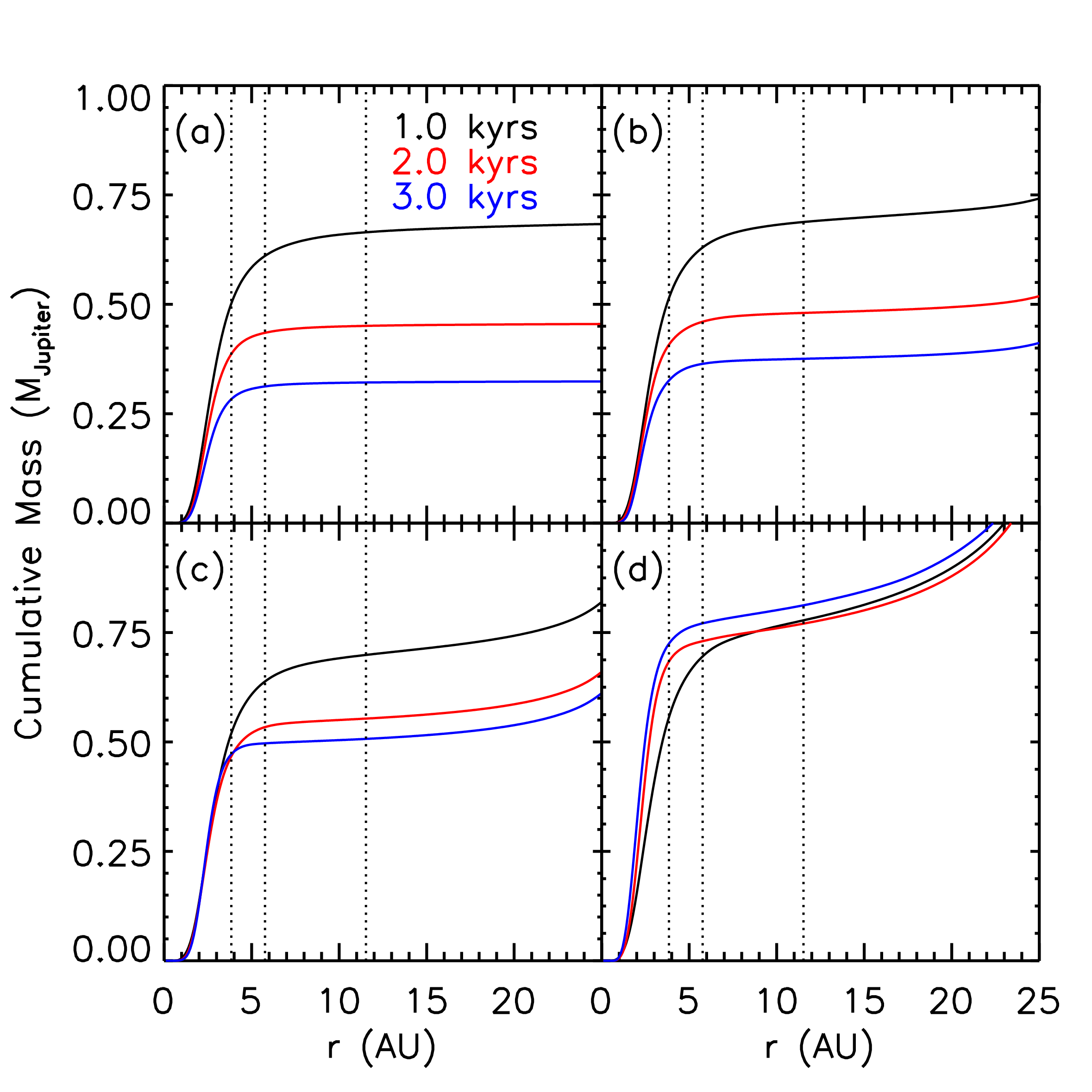}
\caption{Cumulative mass radial profiles at 1.0, 2.0, and 3.0 kyrs in black, red and blue respectively, for simualations (a), (b), (c), and (d) (see Table 2).  
Vertical dotted lines represent the radial location of the Hill radius, 1/2 the Hill radius, and 1/3 the Hill radius.
As the disk evolves, the largest fraction of mass remains within 5 AU. 
This is $\sim1/3$ the Hill sphere of the host planet. 
\label{masshill}}
\end{figure}

\begin{figure}
\includegraphics[width=0.7\textwidth,natwidth=610,natheight=642]{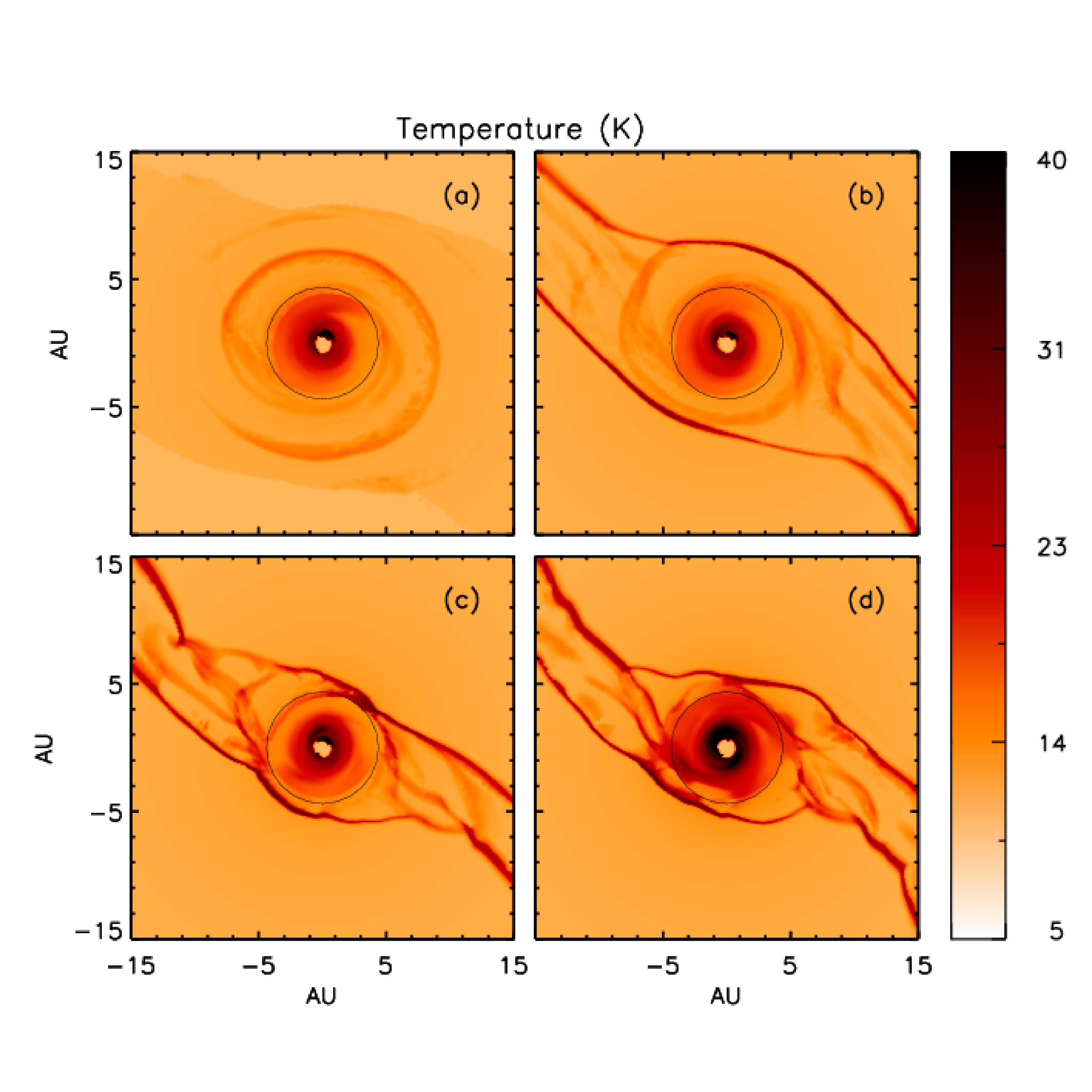}
\centering
\caption{End-state midplane temperature contours (K) of the subdisk at 3 kyrs for simualations (a), (b), (c), and (d) (see Table 2).  
The disk-planet system's Hill radius is $\sim11.53~\mathrm{AU}$ shown as black circles in each panel.  
The disk is truncated at  $R_{Hill}/3$.  
The regions of the disk where material is flowing onto the grid produce high temperature regions that could lead to ice desorption and rotational emission from gas phase molecules.\label{temp_image}}
\end{figure}

\begin{figure}
\includegraphics[width=0.7\textwidth,natwidth=610,natheight=642]{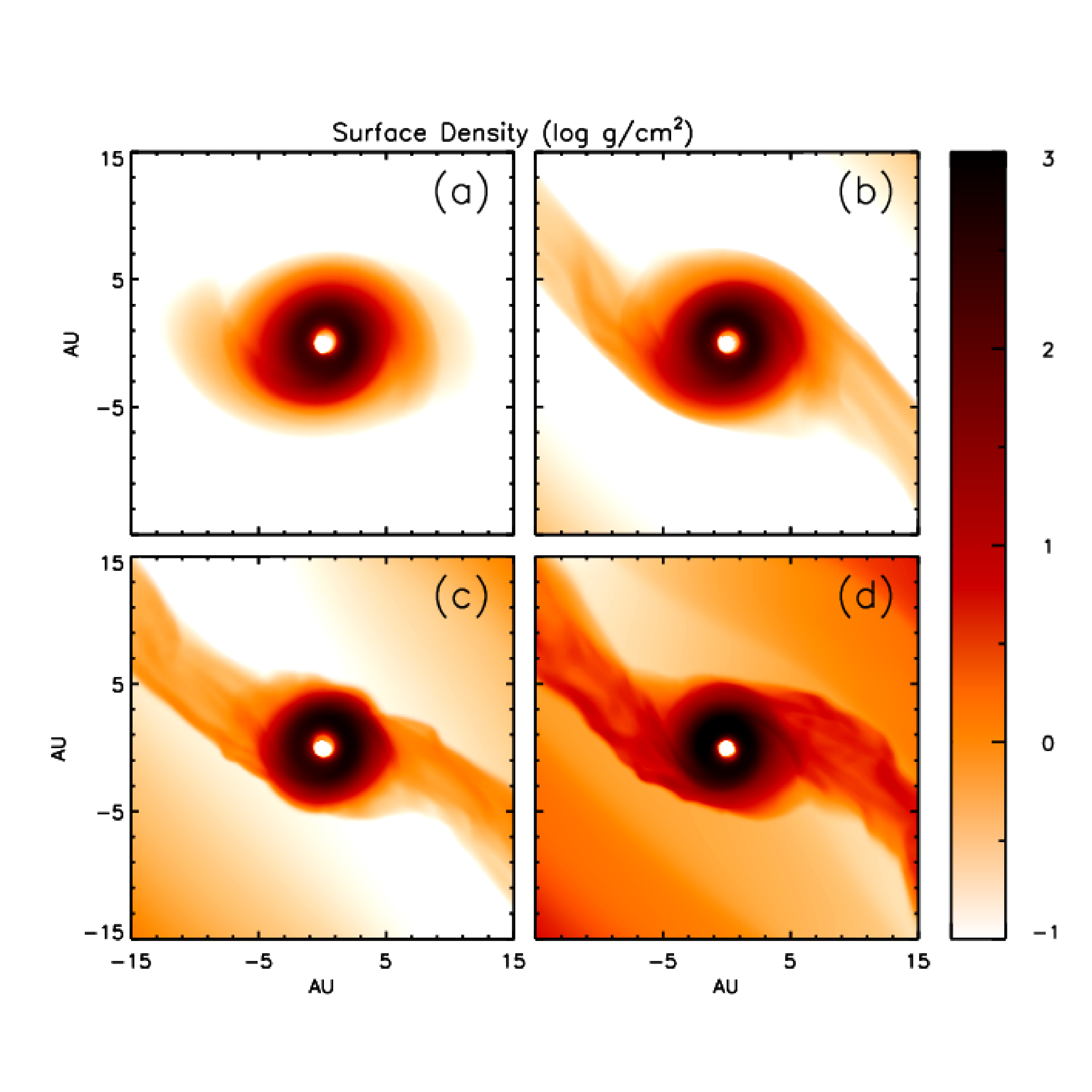}
\centering
\caption{Surface density contour of embedded subdisk at 3 kyrs, in log $\;g/cm^2$, for simualations (a), (b), (c), and (d) (see Table 2).  At radial distances just inside of $\;5 \mathrm{AU}$, the material rapidly increases in density and velocity as the gravitational effects of the host planet start to dominate the flow of mass. The elongated shape of the subdisk is a result of tidal forces from the primary and dynamical interactions with circumplanetary material.\label{endstate}}
\end{figure}

\begin{figure}
\centering
\includegraphics[width=0.6\textwidth,natwidth=610,natheight=642]{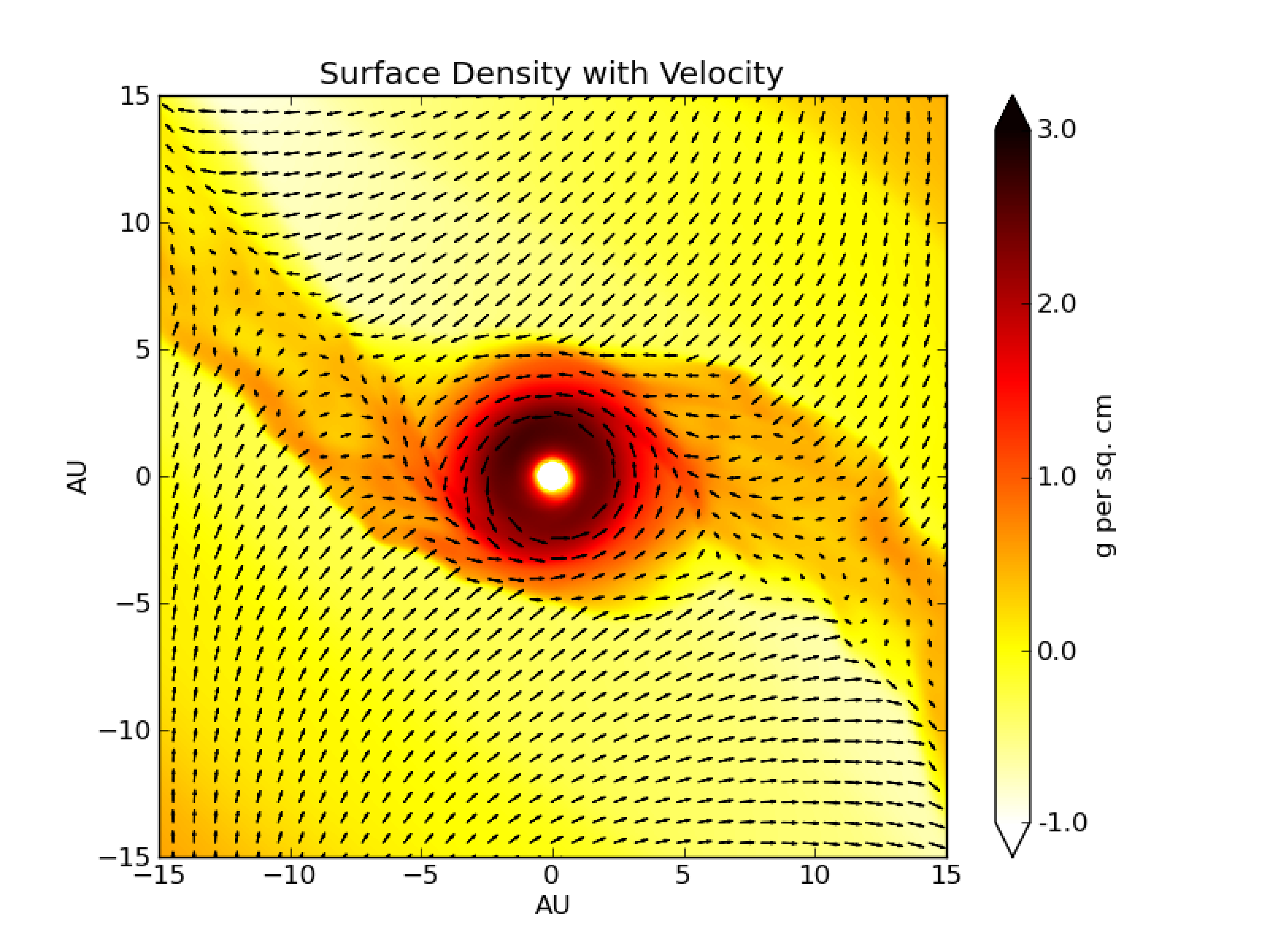}
\centering
\caption{Surface density contour of embedded subdisk at 3 kyrs, in log $\;g/cm^2$, for simualations (d) (see Table 2).  Shown in black are velocity vectors that trace the flow of mass into the subdisk from the surrounding circumstellar disk. \label{endstate_vel}}
\end{figure}

\begin{figure}
\centering
\includegraphics[width=0.7\textwidth,natwidth=610,natheight=642]{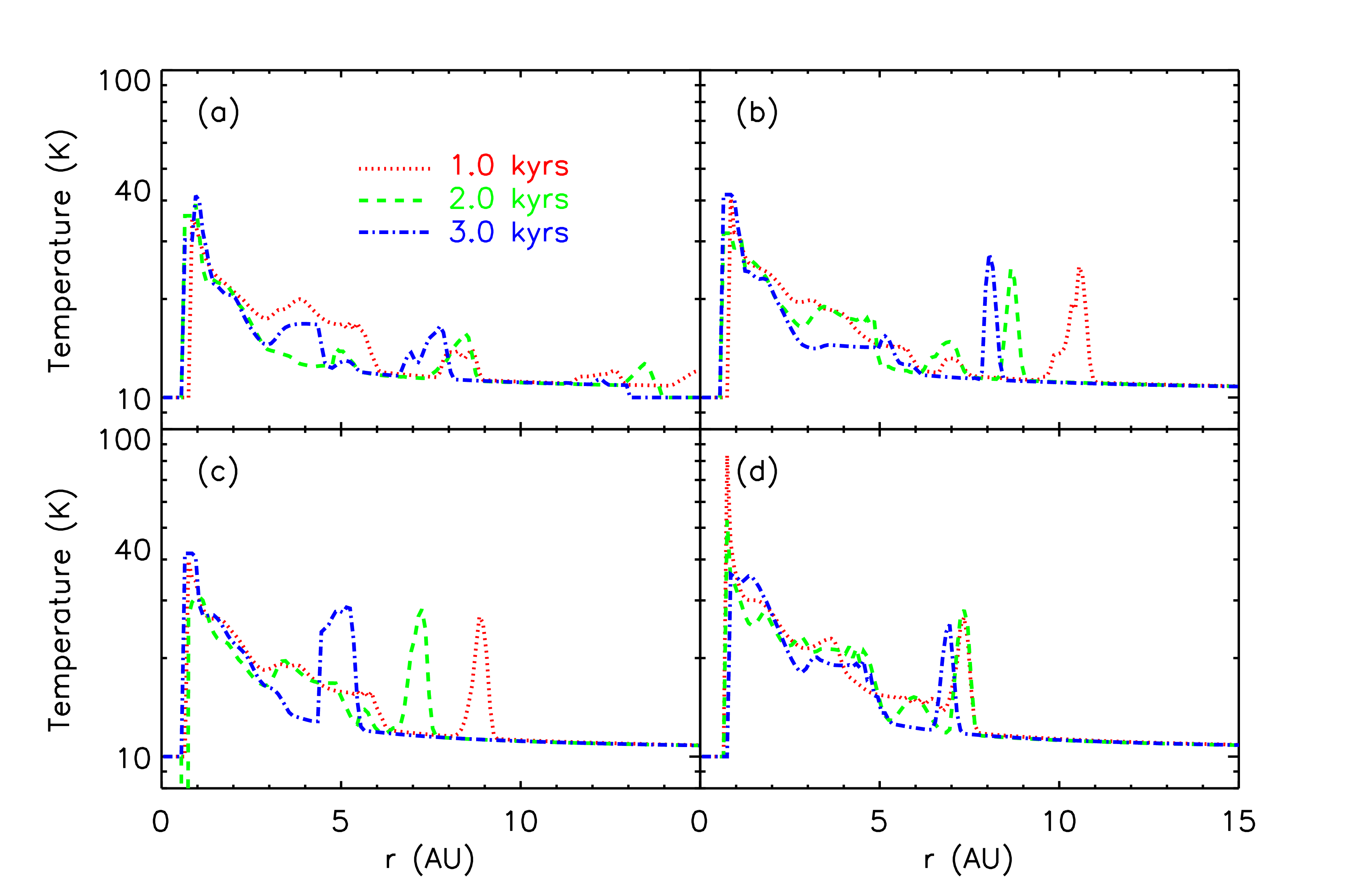}
\caption{Radial cuts of the midplane temperature through the hot molecular wall at three different times, for simualations (a), (b), (c), and (d) (see Table 2). 
The red, green, and blue curves correspond to  1.0, 2.0, and 3.0 kyrs, respectively.  
Shock heating causes temperature variations along spiral arms and where material from the main disk material interacts with the subdisk.  
The inner disk is heated as mass is transferred inward by the spiral arms. 
The temperature variations in the subdisk and at the disk-subdisk boundary can cause desorption of some ices into the gas phase, increasing, e.g., CO abundances.  
Emission from these gas phase molecules may be observable with ALMA.\label{tprof2}}
\end{figure}

\begin{figure*}
\centering
\includegraphics[width=0.4\textwidth,natwidth=610,natheight=642]{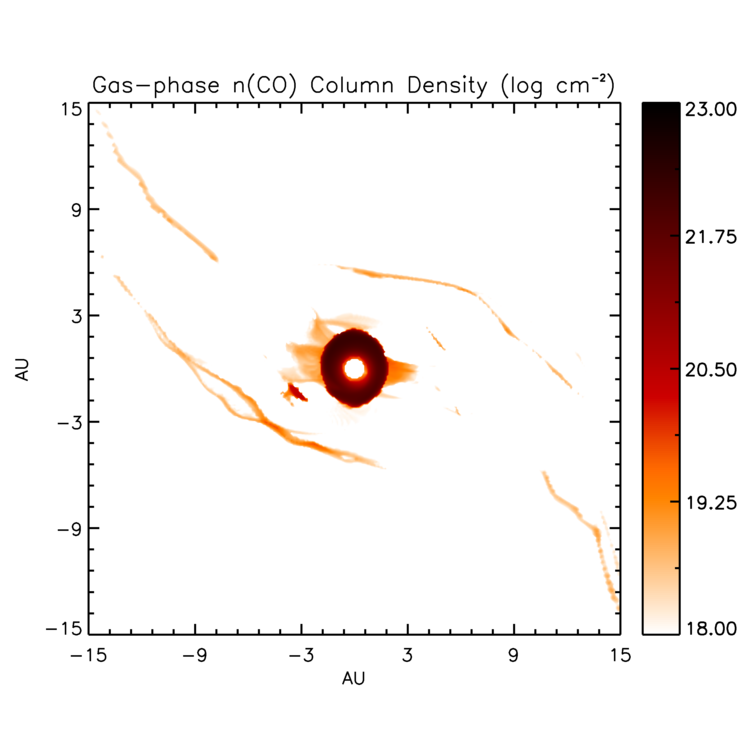}
\includegraphics[width=0.4\textwidth,natwidth=610,natheight=642]{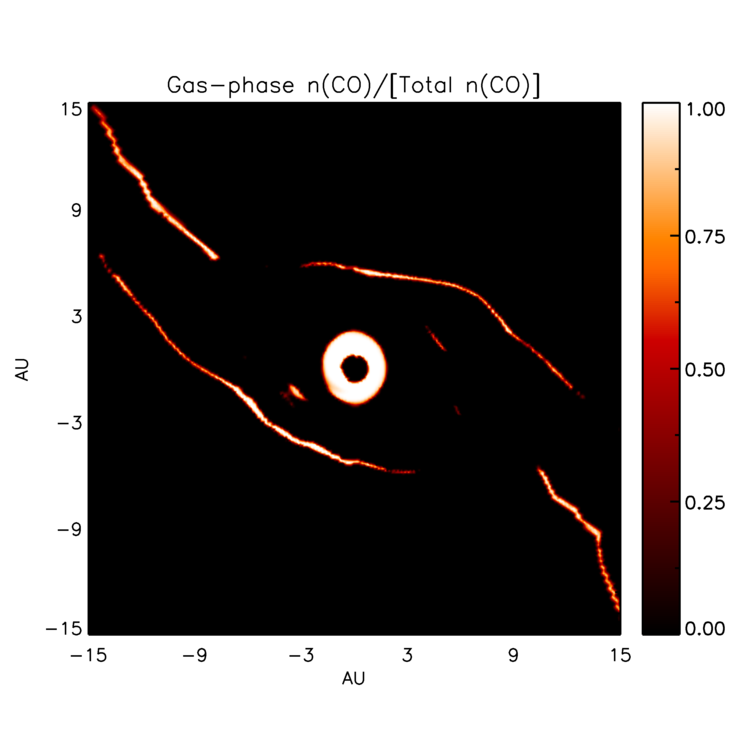}
\caption{Gas-phase CO column densities (left) and the midplane CO number density in the gas phase relative to the total number of CO molecules  (right).  The gas-phase CO is largely confined to the molecular walls and in the warmer regions of the subdisk. Gas-phase CO can have very large column densities, potentially creating strong molecular line emission.  
}
\end{figure*}

\begin{figure}
\centering
\includegraphics[width=0.6\textwidth,natwidth=610,natheight=642]{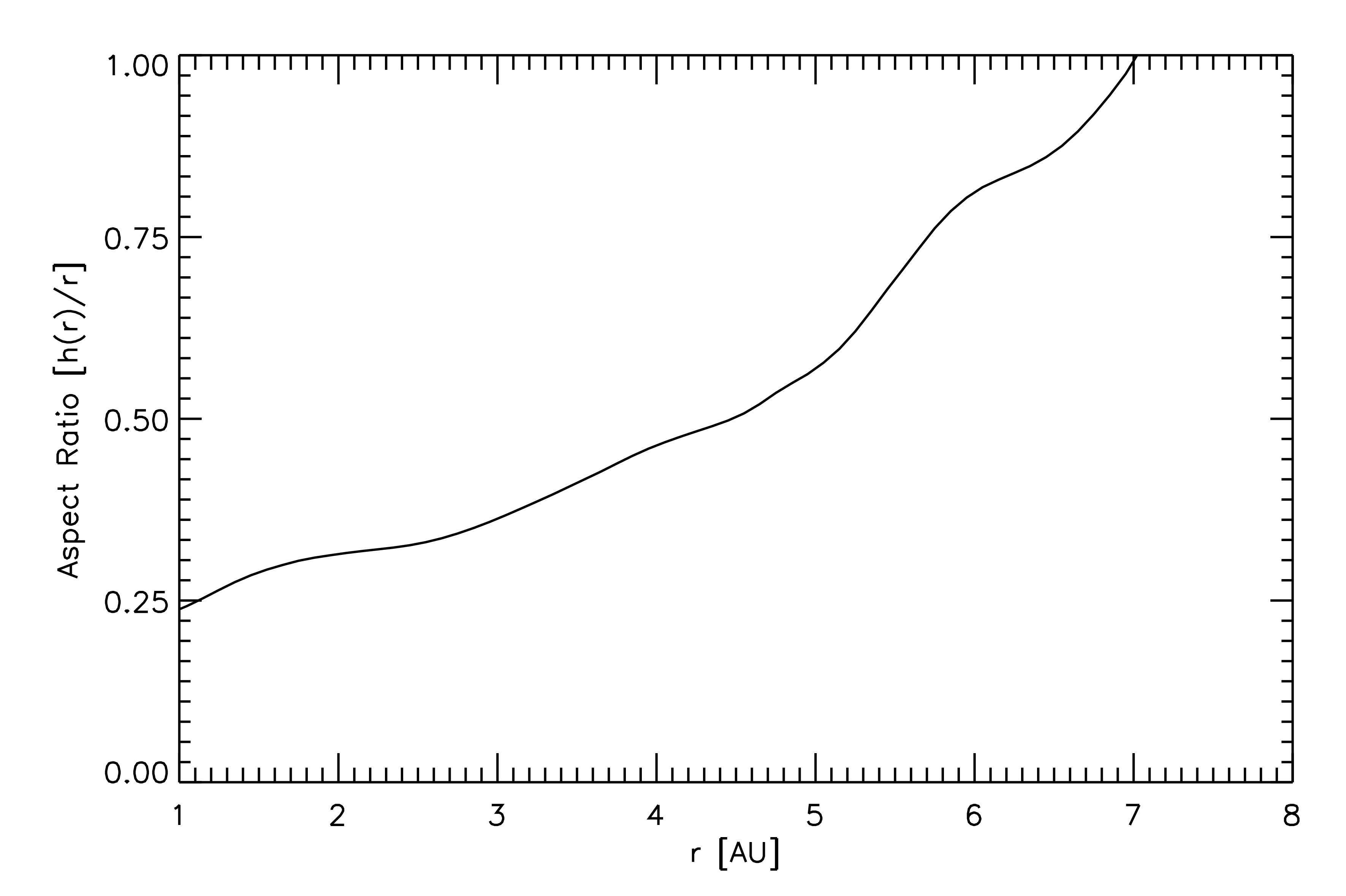}
\caption{azimuthally averaged mass weighted scale height verses radial distance for simulation (d) at 3.0 kyrs.  Subdisks are in the thick disk regime with aspect ratios $(h/r)$ of $\gtrsim0.2$.}
\end{figure}

\begin{figure}
\centering
\includegraphics[width=0.8\textwidth,natwidth=610,natheight=642]{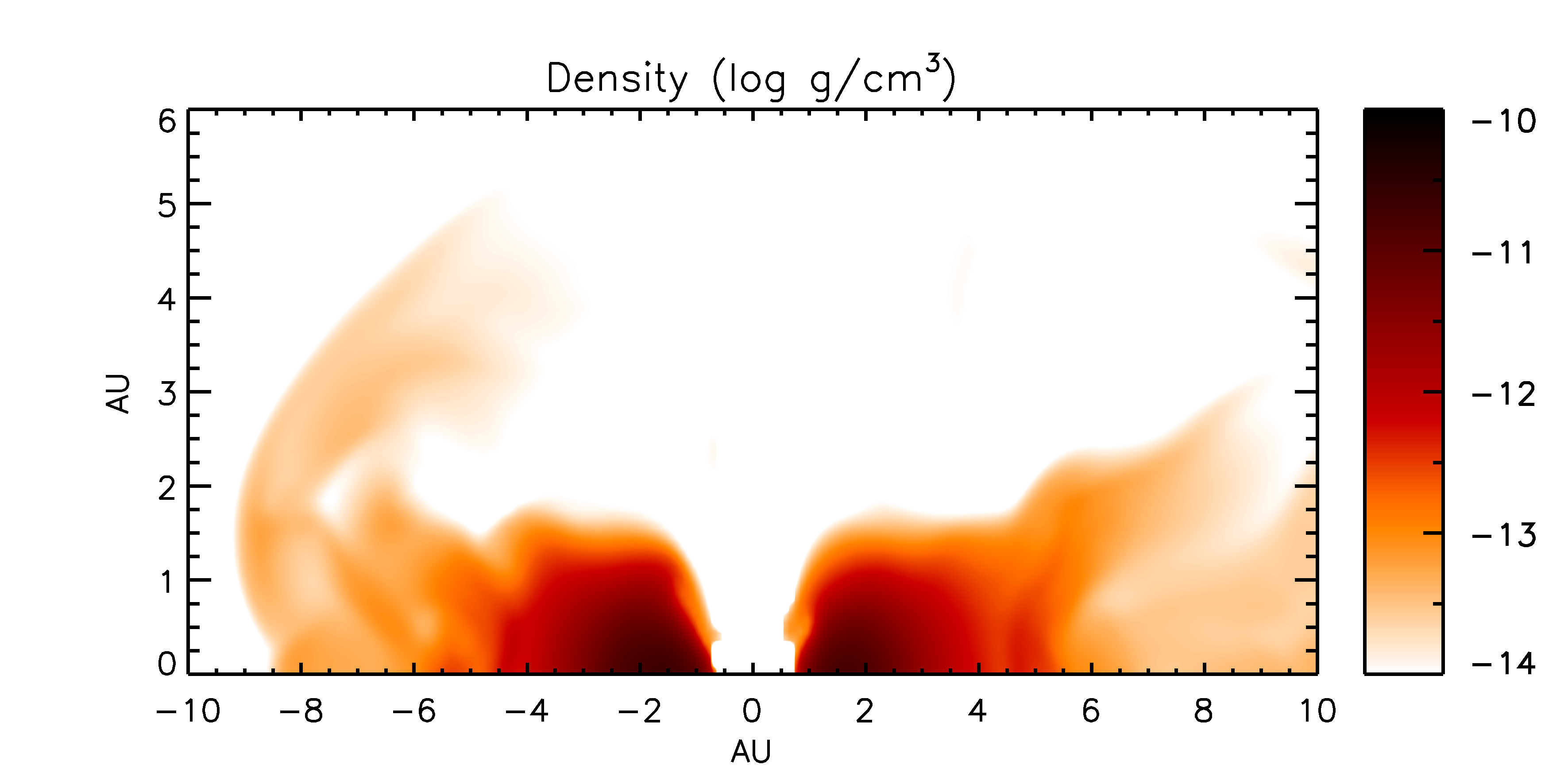}
\caption{Volume density vertical cuts through the substellar and anti stellar regions of the sub disk. The disk aspect ratio shows that circumplanetary disks are in the thick disk regime, which will be important for future moon formation and migration studies.}
\end{figure}

\end{document}